\documentclass[pra,twocolumn]{revtex4-1}
\usepackage{amsmath}
\usepackage{amsfonts}
\usepackage{amssymb}
\usepackage{graphicx}
\usepackage{natbib}

\begin{document}
\title{Self-heterodyne detection of the {\it in-situ} phase of an atomic-SQUID}
\author{R.~Mathew$^{1}$}
\author{A.~Kumar$^{1}$}
\author{S.~Eckel$^{1}$}
\author{F.~Jendrzejewski$^{1}$}
\author{G.~K.~Campbell$^{1}$}
\author{Mark Edwards$^{1, 2}$}
\author{E.~Tiesinga$^{1}$}
\affiliation{$^{1}$Joint Quantum Institute, National Institute of Standards and Technology and University of Maryland, Gaithersburg, Maryland 20899, USA}
\affiliation{$^{2}$Department of Physics, Georgia Southern University, Statesboro, Georgia 30460-8031, USA
}
\begin{abstract}

We present theoretical and experimental analysis of an interferometric
measurement of the {\it in-situ} phase drop across and current flow
through a rotating barrier in a toroidal Bose-Einstein condensate
(BEC).
This experiment is the atomic analog of the rf-superconducting
quantum interference device (SQUID).  The phase drop is extracted from
a spiral-shaped density profile created by the spatial interference of
the expanding toroidal BEC and a reference BEC after release from all
trapping potentials. We characterize the interferometer when it contains
a single particle, which is initially in a coherent superposition of a
torus and reference state, as well as when it contains a many-body state
in the mean-field approximation. The single-particle picture is sufficient
to explain the origin of the spirals, to relate the phase-drop across the barrier to
the geometry of a spiral, and to bound the expansion times for which the
{\it in-situ} phase can be accurately determined.  Mean-field estimates and
numerical simulations show that the inter-atomic interactions shorten the
expansion time scales compared to the single-particle case. Finally, we compare
the mean-field simulations with our experimental data and confirm that
the interferometer indeed accurately measures the {\it in-situ} phase drop.

\end{abstract}
\maketitle
\section{Introduction}

Atomtronics focuses on the creation of atomic analogues to electronic
devices. Analogues to several electronic components, such as diodes
and transistors, have been proposed~\cite{seaman_2007}, while several
other circuit elements have been experimentally realized, including
capacitors~\cite{Lee2013, krinner_2015} and spin-transistors~\cite{Beeler2013}.
The atomic version of the rf-superconducting
quantum interference device (SQUID) has been realized
\cite{Ramanathan2011,wright_2013,Eckel2014}, and initial experiments
towards the creation of a dc-SQUID have been performed
\cite{Ryu2013,Jendrzejewski2014}. Both SQUID devices are formed using
a toroidal Bose-Einstein condensate and contain one or more rotating
weak links or barriers. Furthermore, creation of an atomic 
rf-SQUID in a ring-shaped lattice 
has been proposed~\cite{amico_superfluid_2014,aghamalyan_coherent_2015}.
Theoretically persistent current states in (quasi) one-dimensional
toroidal geometry have been studied extensively
\cite{kashurnikov_1996,kagan_2000,buchler_2001,cominotti_2014}.  Weak
links, whether superconducting or atomic, are characterized by the
relationship between the current through and the phase across the
barrier~\cite{Likharex1979}.  Accurate measurement of this current-
phase relationship in the atomic system is crucial for the
characterization of atomtronic devices.

Measurement of the {\it in-situ} phase of a condensate through
interference is a common tool in modern cold-atom physics.
Since the first interference between three-dimensional condensates
was demonstrated in 1997~\cite{Andrews1997,Castin1997}, several
experiments have used interference to infer details about the {\it
in-situ} phase profile of condensates \cite{simsarian_imaging_2000}. 
Vortices in condensates~\cite{inouye_2001} and fluctuations brought on
by the two-dimensional Berezinskii-Kosterlitz-Thouless phase
transition~\cite{Hadzibabic2006} have also been detected interferometrically.
Interference between two molecular BECs \cite{Kohstall2011} and BECs on
an atom chip have also been observed \cite{Shin2005}. 
Recently, interference measurements have been extended to determine the
persistent current state in a toroidal condensate~\cite{Corman2014,Eckel2014b}.
Reference~\cite{Eckel2014b} also measured the current-phase relationship
of a BEC in a toroidal trap with a rotating barrier, the atomic analogue
of an rf-SQUID.  

\begin{figure}
    \includegraphics[scale=1]{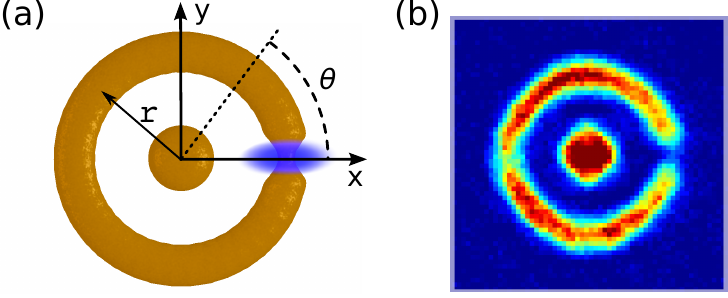}
    \caption{ Panel (a) shows a schematic of atoms in a target trap.
    The inner disc and the outer ring are the reference and science condensates,
    respectively. A blue-detuned laser forms a rotating weak link
    and is shown by the blue ellipse.
    Panel (b) is an {\it in-situ} image from the experiment of atoms in a target trap.
   } 
   \label{fig:exp_setup}
\end{figure}

In the experiment of Ref.~\cite{Eckel2014b} a single condensate was
created in a simply connected trap and subsequently split into two
condensates. One condensate was confined in a toroidally shaped
``science'' trap and the other condensate was confined in a concentric
disc-shaped ``reference'' trap.  We will refer to these together as
the ``target'' trap. A schematic and an {\it in-situ} image of
atoms in a target trap is shown Fig.~\ref{fig:exp_setup}. The
science and reference traps were separated by more than 5~$\mu$m, thus
atom tunneling between them is negligible and the condensates dephase
rapidly because of imperfections in the splitting procedure. Hence,
when the two condensates expand and interfere after turning off all
trapping potentials, their relative phase is random, thus representing
a self-heterodyne measurement \cite{Richter1986}.  Rotating weak links
are only applied to the condensate in the science trap and the other
condensate is a phase reference.  The current through and the phase
drop across the barrier were inferred from the spiral-shaped
modulation in the density profile for short expansion times. The
number of spiral arms determines the winding number of the persistent
current state, while their chirality determines the direction of atom
flow.

In this paper, we study in detail the interference patterns that result
from interfering a toroidal condensate with a reference condensate and
verify the interferometric technique used in Ref.~\cite{Eckel2014b}
to measure the current-phase relationship. 
We first study analytically and numerically a single-particle version 
of the atomic rf-SQUID
in section \ref{sec:single_particle}.  We find that the experimentally
observed spirals are a short time phenomenon and both the current through
and phase drop across the barrier follow from the geometry of the
spirals. For longer expansion times, the spirals become modulated with
concentric circles due to self-interference of the torus and
it becomes difficult to read out
the {\it in-situ} phase drop.  
In Sec.~\ref{sec:int} we describe details of our experiments with sodium
condensates in a target trap.  In addition, this section describes the
numerical techniques used to simulate the mean-field Gross-Pitaevskii
equation, which quantifies the effects of atom-atom interactions on the
expanding condensates.  Estimates of bounds on expansion times, where
spirals can be observed in the density profile, are also derived.
Finally, a comparison of theoretical and experimental results in 
Sec.~\ref{sec:expt} validates the interferometric method 
for measurement of the current-phase relationship of an atom-SQUID.

\section{Single-particle picture}
\label{sec:single_particle}
We begin our study of the interference by deriving analytic expressions
for the free expansion of a single atom of mass $m$ released from a
target-trap interferometer and give an intuitive explanation of the
origin of the spirals in the interference pattern. 
To generate the interference, we assume that the wavefunction of our
single particle is in a superposition of a wave localized in the reference
and science regions, respectively. 

\subsection{Particle in a rotating torus}
In order to solve for the wavefunctions, we first describe the target
trap in cylindrical coordinates $\vec x=(r,\theta,z)$.  The science
and reference traps are assumed to be parabolic in the radial
direction, and centered at $r_S$ and the origin, respectively
(see Fig.~\ref{fig:exp_setup}(a)). The harmonic
oscillator lengths are $\sigma_S$ and $\sigma_R$, respectively. The
common transverse confinement is harmonic with oscillator length
$\ell_z$. We assume $\sigma_R \approx \sigma_S$ and $\sigma_S,
\sigma_R, \ell_z\ll r_S$. In addition, the science trap has a barrier
or weak link rotating at angular frequency $\Omega$ inducing atom
flow.  For simplicity, we model the barrier in the science trap as a
Dirac delta-function $V_b(\vec x,t) = U_0 w(r)\delta(\theta -\Omega
t)$ with strength $U_0$, time $t$, and $w(r)$ is a window function
which is one around the radial position of the science trap and zero
everywhere else.

In the frame rotating with the barrier the atom is
prepared in the time-independent state $\Psi_{\rm init}(\vec
x)=(\psi_R(\vec x)+\psi_S(\vec x))/\sqrt{2}$, where the $\psi_i(\vec
x)=\eta_i(r)\varphi_i(\theta)\phi_z(z)$ are separable wave functions
of the science ($i = S$) and reference ($i=R$) trap. Here, $\phi_z(z)$
is the unit-normalized 1D ground state harmonic-oscillator wavefunction
and $\eta_i(r)= e^{-(r-r_i)^2/(2\sigma_i^2)}/{\cal N}$ is the radial
wavefunction, where ${\cal N}$ is a normalization constant.  The overlap
between the $\psi_i(\vec x)$ is negligible.

\begin{figure}
    \includegraphics[scale=0.32]{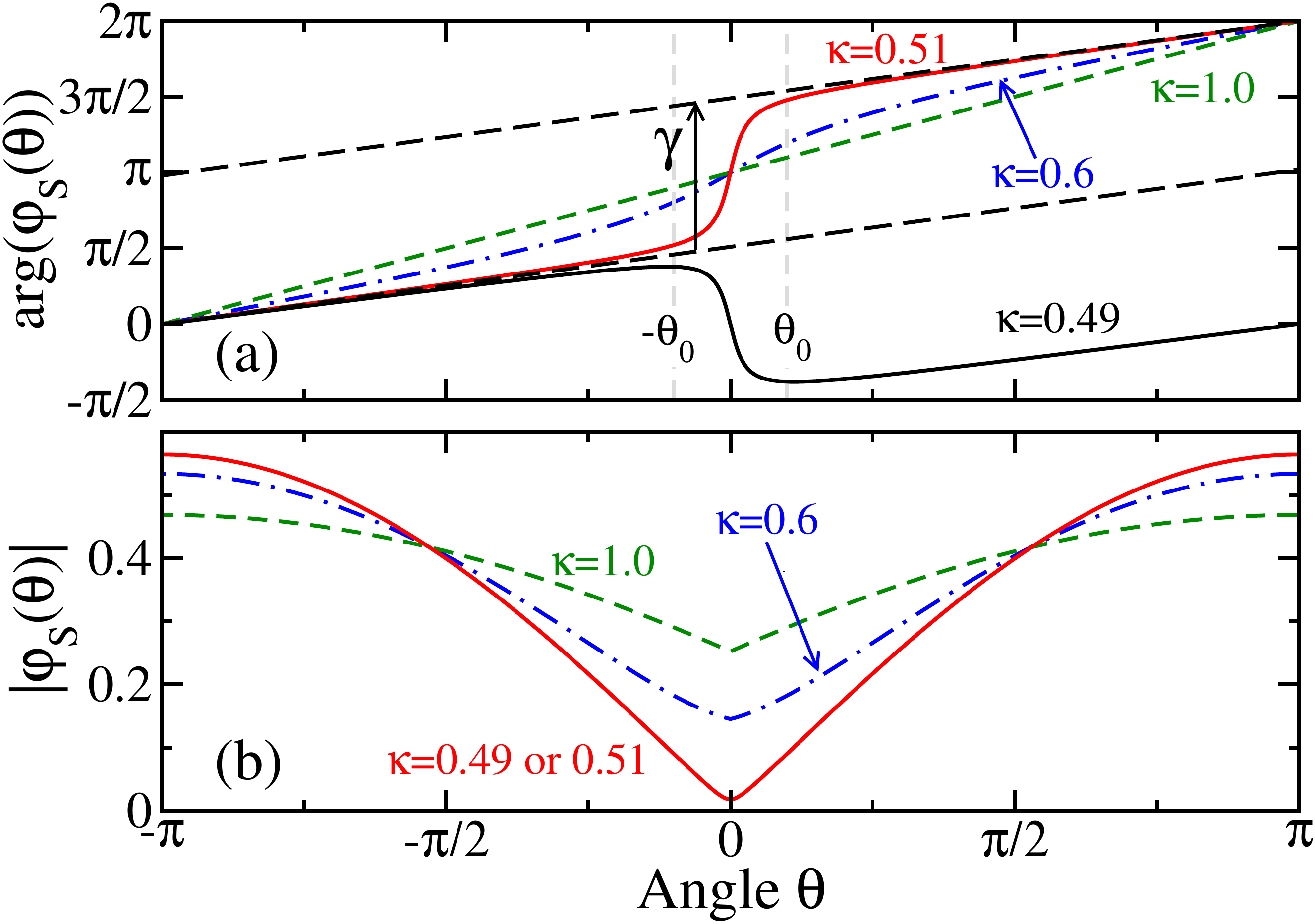}
    \caption{
    Phase (panel a) and magnitude (panel b) of the single-particle ground-state angular
    wavefunction $\varphi_S(\theta)$ as a function of $\theta$ for various
    values of rotation rate $\kappa$. The wavefunction is calculated in the frame rotating 
    with a delta-function potential
    of strength $U_0=1$ located at $\theta=0$.
    For $\kappa \approx 1/2$, a sharp change
    in the phase occurs in the $\kappa$-dependent region $\theta \in (-\theta_0,
    \theta_0)$. The figure also shows the phase-drop $\gamma$, defined in the text,
    for $\kappa = 0.51$.
   } 
   \label{fig:phase}
\end{figure}

The angular functions $\varphi_i(\theta)$ are $1/\sqrt{2\pi}$
for the reference trap and the ground state of the Schr\"odinger
equation $[-d^2/d\theta^2+2i\kappa d/d\theta +U\delta(\theta)]
\varphi_S(\theta)=E\varphi_S(\theta)$ for the toroidal trap with
rotating barrier. Here, $\kappa= \Omega/\Omega_0$,
 $U=U_0/{\cal E}_0$, $\Omega_0=2{\cal E}_0/\hbar$, and ${\cal E}_0=\langle\hbar^2/(2m
r^2)\rangle\approx\hbar^2/(2m r_S^2)$ is the natural energy scale of
the science trap, where the bracket $\langle\cdot\rangle$ indicates an
expectation value over $r$ and $z$ and $\hbar$ is the reduced Planck's constant. The
function $\varphi_S(\theta)$ is periodic on $\theta\in[-\pi, \pi]$ and
a superposition of $\exp[i(\kappa\pm\sqrt{E + \kappa^2})\theta]$ with
energy $E = -\kappa^2 + \epsilon(\kappa)$, where $\epsilon(\kappa)$
is periodic in $\kappa$ with period one.  Examples of the phase and
magnitude of $\varphi_S(\theta)$ are shown in Fig.~\ref{fig:phase}. For
most $\kappa$ the phase of $\varphi_S(\theta)$ changes nearly linearly
with $\theta$. Only for $\kappa \approx 1/2$ and, in fact, near any
half-integer $\kappa$ it changes rapidly near the barrier at
$\theta =0$. This rapid change around $\theta \in (-\theta_0, \theta_0)$
is accompanied by a decrease in density.  The phase jump is $\pi(-\pi)$
for $\kappa$ just above (below) $1/2$ and the density is zero at $\theta
= 0$ for $\kappa=1/2$. 

We define the phase drop across the barrier as $\gamma = 2\pi (n - s)$,  
where $n$ is the winding number, which for the ground-state of the single-particle
wavefunction equals to the integer closest to $\kappa$, and the slope
\begin{equation}
    s = \left.\frac{d}{d\theta}\left[ \arg(\varphi(\theta)) \right]\right\vert_{\theta=-\pi}.
    \label{eq:slope}
\end{equation}
A graphical representation of $\gamma$ for $\kappa = 0.51$ is shown in Fig.~\ref{fig:phase}. 
In the rotating frame, the angular current 
\begin{equation}
    \label{eq:J}
J(\kappa)
= r_S\Omega_0|\varphi_S(\theta)|^2\left( 
        \frac{d\arg[\varphi_S(\theta)]}{d\theta} - \kappa \right),
\end{equation} 
for any $\theta$ and we used the fact that $\langle 1/r \rangle = 1/r_S$.

\subsection{Single-particle interference}
After turning off the target trap the atomic wavefunction , $\Psi(\vec x,t)$,
freely expands and interferes. At time $t$ after the release, it is imaged
along the $z$ axis leading to the observable
$
     n(r,\theta,t) = \int_{-\infty}^\infty dz\, |\Psi(\vec x,t)|^2,
$
where $\Psi(\vec x, t=0) = \Psi_{\rm init}(\vec x)$.
During the expansion, the wavefunction of the torus and disc remains separable in the 
$z$ direction, i.e.  $\psi_i(\vec x,t) = \chi_i(r, \theta, t)\phi_z(z, t)$.
Thus, $n(r,\theta,t) = |\chi_R(r, \theta, t) + \chi_S(r, \theta, t)|^2$ as 
$\int dz |\phi_z(z,t)|^2 = 1$. 

\begin{figure}
\includegraphics[width=\columnwidth]{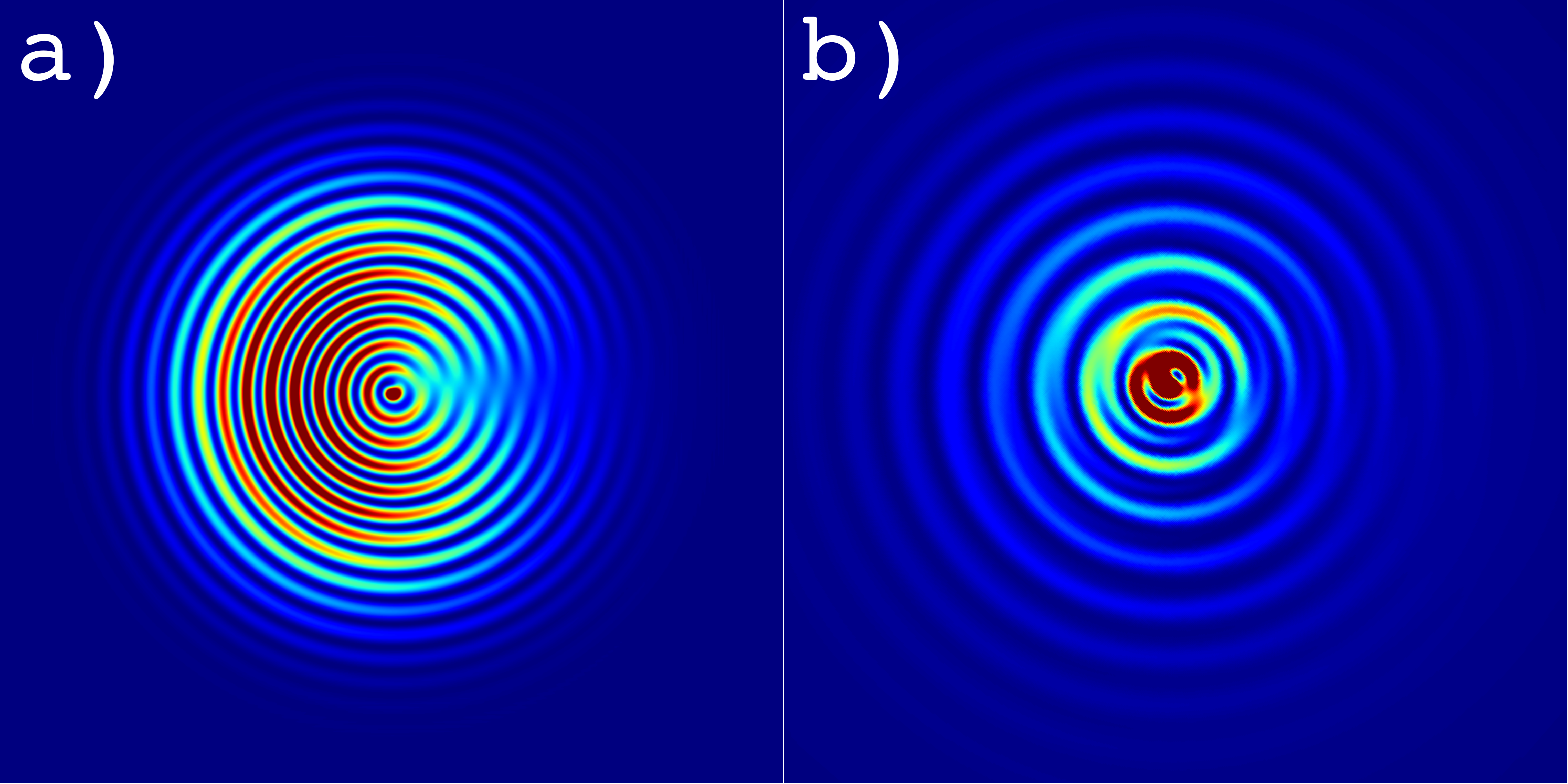}
        \caption{
Numerical simulation of the integrated particle density $n(r,\theta, t)$
of a single particle, with winding number equal to one, 
expanding in the rotating frame after release from a
target trap.  Panel (a) shows $n(r, \theta, t)$ with spirals at an early
expansion time $t=0.25\tau_C$, evaluated at $r=r_S$. Panel (b) shows
a later time $t = 1.25\tau_C$, where the spirals are superimposed with
circles due to self-interference of the toroidal wavefunction.  The density
near the center has been truncated for better contrast.  The trap
parameters are $ \sigma_R = 0.025 r_S,\, \sigma_S = 0.05 r_S, \, U_0 =
1$ and $\kappa = 0.51$. The length of the sides in panels (a) and (b)
corresponds to $5.12 r_S$ and $12.8 r_S$, respectively. The parameters are
chosen such that the overlap between the expanding science and reference
wavefunctions is sufficient to show the spiral over a large range of radii.
} 
\label{fig:1p_inter} 
\end{figure}

It is convenient to first follow the expansion with a numerical solution
of the Schr\"odinger equation in the $(r, \theta)$ plane for $\kappa$ near
$1/2$. Figure \ref{fig:1p_inter} shows $n(r, \theta, t)$ for two different
expansion times. (Time propagation was carried out by switching to momentum
space, applying appropriate time-dependent phase factors, and returning
back to coordinate space.) We observe that as soon as the wavefunctions
of the two traps overlap, the interference pattern consists of
spirals. Later on, the self-interference of the science wavefunction yields
circles superimposed on the spirals. 

We confirm these interference patterns with an
asymptotic expansion and study the associated time-scales.
The time evolution of the reference state is
\begin{equation}
\chi_R(r, \theta, t) = e^{-r^2/2\sigma^2_R(t)}/{\cal N}_1(t),
\label{eq:ref_wf}
\end{equation}
where $\sigma_R^2(t) = \sigma_R^2 + i\hbar t/m$
and ${\cal N}_1(t)$ normalizes the wavefunction. Hence, for $t \gg
m\sigma_R^2/\hbar $ the spatial extent of the reference wavefunction,
$\sqrt{|\sigma_R^2(t)|}$, is proportional to the expansion time,
corresponding to ballistic expansion.  In contrast, the expanding science
wavefunction is not analytically solvable. We can, however, derive an
asymptotic series based on the pertinent timescales of the expansion of
the science wavefunction.  The shortest time scale is the ballistic time
$\tau_B = m\sigma_S^2/\hbar$ determined by the initial radial width. In
addition, as will become clear later, 
there are two position-dependent
timescales: an intermediate timescale $\tau_C(r) = m\sigma_S(r+r_S)/\hbar$
and a long timescale $\tau_S(r) = m r r_S/\hbar$.  We are interested
in the expansion time interval $\tau_B \ll t \ll \tau_S(r)$.  Figure
\ref{fig:1p_inter} shows the density profile for two such times.

Formally, expanding wavefunction $\chi_S(r, \theta, t)$ evolves as 
\begin{align}
    \label{eq:Evol}
    \chi_S(r,\theta, t) &= \int_0^\infty dr'\, r' \eta_S(r') \nonumber \\ 
    & \times\int_{-\pi}^{\pi} d\theta'\,
                    G(r, \theta, r', \theta', t) \varphi_S(\theta'),
\end{align}
where the free-particle Green's function \cite{Feynman_QM} in two-dimensions 
is
\begin{align}
    \label{eq:Green_fn}
G(r, \theta, r', \theta', t)&=
\frac{m}{2\pi i \hbar t} \times \\
&\exp\left\{ \frac{im\left[ r^2 + r'^2 - 2 r r' \cos(\theta -\theta') 
            \right]}{2\hbar t} \right\}\nonumber.
\end{align}
We note that the integral over $r'$ is concentrated around $r'=r_S$.
Consequently, the integral over $\theta'$ in Eq.~\ref{eq:Evol} can be
solved by noting that the phase on the right hand side (RHS) of Eq.~\ref{eq:Green_fn}
oscillates rapidly for $t \ll mrr'/\hbar\approx \tau_S(r)$.  Then, the
method of steepest-descent \cite{Bender_Orszag} gives an asymptotic series
for integral over $\theta'$ in powers of the small parameter $t/\tau_S(r)$.
In fact, there are two stationary points located at $\theta' = \theta$
and $\theta' = \theta+\pi$, respectively.  The remaining integral over $r'$ is also
solved using steepest descent for $\tau_B \ll t$
based on the small parameter $\sigma_S/r_S$.
To leading order we find
\begin{align}
    \label{eq:psi_lead} 
    \chi_S(r, \theta, t) =& \left[e^{-(r-r_S)^2/[2\sigma^2_S(t)]}
        \varphi_S(\theta)\right. \\
      &\left.+ \, e^{-(r+r_S)^2/[2\sigma^2_S(t)]}
            \varphi_S(\theta + \pi)\right]/
      ({\cal N}_2(t)\sqrt{r})\nonumber
   \, ,
\end{align}
where the complex, time-dependent $\sigma_S^2(t) = \sigma_S^2(1 + i
t/\tau_B)$ is the square of the width of the expanding radial wave-packet
and $1/{\cal N}_2(t)$ is a normalization factor.  The wavefunction is
a superposition of two expanding 1D Gaussians centered at $r_S$ and
$-r_S$ (except for the probability conserving factor $1/\sqrt{r}$).
The asymptotic solution is valid for $\tau_B \ll t \ll \tau_S(r)$. This
excludes the region near the origin, where $\tau_S(r)$ is small.

It is natural to ask whether the second term in Eq.~\ref{eq:psi_lead}
is important relative to the first term. Clearly, when
$\sqrt{|\sigma^2_S(t)|} < r+r_S$ or equivalently $ t <\tau_B
(r +r_S)/\sigma_S =\tau_C(r)$ the second term is negligible.
The interference of the first term with the reference wavefunction
$\chi_R(r, \theta, t)$ in Eq.~\ref{eq:psi_lead} leads to spirals in the
density $n(r, \theta, t)$ as shown in Fig.~\ref{fig:1p_inter}(a).  For $t
\ge \tau_C(r)$ the second term cannot be ignored and interferes with the
first term. It leads to circles in addition to the spirals as shown in
Fig.~\ref{fig:1p_inter}(b). An intuitive interpretation of $\tau_C(r)
$ is that it corresponds to the time taken by signals from both antipodal points $(r_S,
\theta)$ and $(r_S, \theta+\pi)$ of the initial $t=0$ wavefunction of the torus to
reach the observation point $(r, \theta)$ and interfere. This is the
self-interference of the toroidal wavefunction.

\subsection{Spirals}
We are now in a position to quantify the spiral structure for $\tau_B \ll t \ll \tau_C$. 
We write
$\chi_i(r,\theta, t) = \sqrt{n_i(r,\theta, t)} \exp[i \xi_i(r, \theta,
t)]$, where $n_i(r, \theta, t)$ is the probability density and $\xi_i(r,
\theta, t)$ is the phase.  The integrated density becomes
\[
    n(r, \theta) = n_S(r, \theta) + n_R(r, \theta)
    + 2 \sqrt{n_S(r, \theta) n_R(r, \theta)}\cos\xi(r, \theta),
\]
where $\xi(r, \theta) = \xi_S(r, \theta) - \xi_R(r)$ and we 
suppress the time argument for notational simplicity.
The last term on the RHS of this equation describes the interference of
the wavefunctions in the two traps.

For the above time interval the second term in Eq.~\ref{eq:psi_lead}
can be ignored, so that $n_R(r, \theta, t)$ is independent of
$\theta$, $n_S(r, \theta, t)$ is a separable function of $r$ and
$\theta$, and  $\xi(r, \theta) \approx \arg[\varphi_S(\theta)]- \hbar
r r_S/(mt)$. (The argument $\arg[\phi_S(\theta)]$ is defined as a
monotonic function of $\theta$.)  Then, spirals correspond to curves
of constant phase $\xi(r, \theta)$ in the $(r, \theta)$ plane.  The
densities $n_i(r, \theta)$ only lead to a slowly-varying envelope in
$r$ and suppression of the signal near $\theta = 0$ that is most
pronounced for half-integer $\kappa$.  Consequently, a spiral is described by 
the parametric curve 
$r(u) = (\xi_0 + \arg[\varphi_S(\theta(u))])\times \hbar t/(m r_S)$ and 
$\theta(u) = -\pi + u \mod 2\pi$, where
$\xi_0$ is a constant (typically chosen such that $n(r, \theta)$ 
is a local extremum) and $u$ is the 
free parameter.  
In the absence of a rotating barrier but for a non-zero 
winding number $n$ of the toroidal state, we find $\arg[\varphi_S(\theta)] =
n\theta$ and the interference pattern has Archimedean spirals with
$r(u) = (\xi_0 + n u )\times \hbar t/(m r_S) $ and $\theta(u) = -\pi +
u \mod 2\pi$. These smooth spirals have been observed experimentally 
\cite{Eckel2014b,Corman2014}.

\begin{figure}
  \includegraphics[scale=0.25]{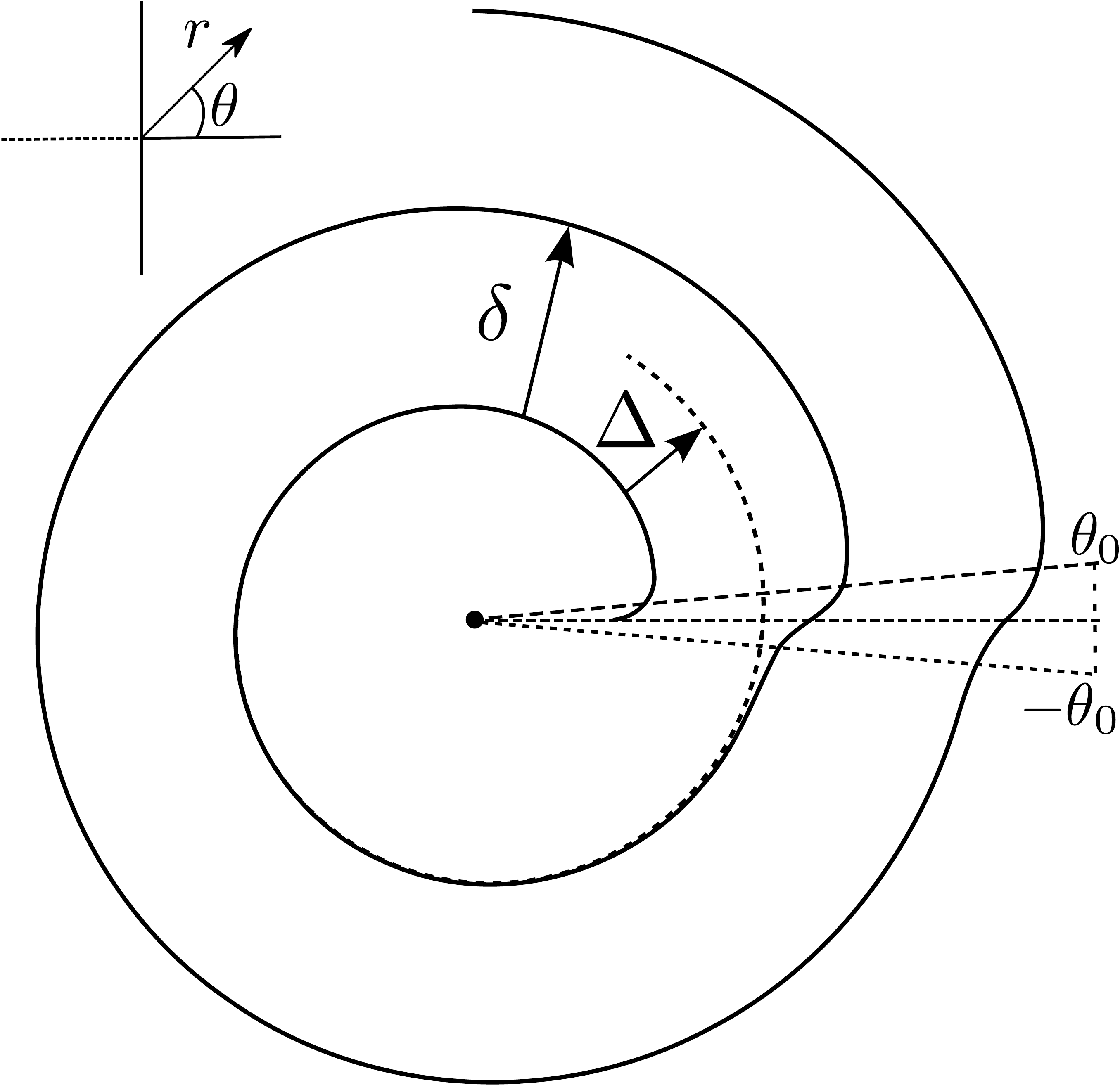} 
  \caption{
Schematic of a spiral-like contour (solid line) in the integrated density
for $\kappa$ slightly larger than $1/2$, so that the winding number $n = 1$. 
The contour has a constant
phase $\xi(r, \theta) =\xi_0$.  The phase of $\varphi_S(\theta)$ varies
rapidly in the wedge $\theta \in (-\theta_0, \theta_0)$. In addition,
a Archimedean spiral (dashed line) with the same initial angular velocity
as the solid line is shown. Its parameters as well as the lengths $\delta$
and $\Delta$ are defined in the text.
}
  \label{fig:spirals}
\end{figure}

A schematic of a spiral is shown in Fig.~\ref{fig:spirals} at a single
expansion time $t$ for $\kappa$ slightly greater than $1/2$, a case
where $\varphi_S(\theta)$ has a sharp phase jump across the barrier near
$\theta =0$. For $|\theta| > \theta_0$ the spirals smoothly wind around
the origin. In contrast, for $\theta \in ( -\theta_0, \theta_0)$ there
is a sharp, nearly discontinuous change in the spirals.  For $\kappa$
away from half-integer values the spirals are smooth everywhere.
The geometry of a spiral is completely determined by the phase $\xi(r,
\theta)$ where the number of spiral arms is the winding number $n$.
The densities $n_R(r, \theta)$ and $n_S(r, \theta)$ determine how many
windings of a spiral are visible along the radial direction.

We characterize the discontinuity or jump of the spirals by lengths
$\delta$ and $\Delta$ shown in Fig.~\ref{fig:spirals}. The quantity
$\delta = 2\pi\hbar t/ (mr_S)$ 
is the radial fringe spacing and measures the increment in $r$ 
as $\xi(r, \theta)$ is increased by $2\pi$ at a fixed $\theta$. 
Moreover, $\Delta = r_A(u
+ 2\pi) - r_A(u) = s \times 2\pi \hbar t/ (m r_S)$, where we used the
Archimedean spiral $r_A(u) = (\xi_0 + s u)\times \hbar t/(m
r_S)$ and $\theta_A(u) = -\pi + u \mod 2\pi$, and $s$ is defined by
Eq.~\ref{eq:slope}. Intuitively, $\Delta$ is the radial distance covered
by a spiral when it is smoothly continued across the barrier region.
The two lengths depend on the dimensions of the torus and expansion
time $t$.

The ratio $\Delta/\delta = s$ is independent of the radial wavefunction
and expansion time.  In fact, we can interpret $\Delta/\delta$
as a measurement of the phase across the barrier $\gamma$, since
\begin{equation}
    \label{eq:phase_drop}
\gamma = 2\pi(n - \Delta/\delta).
\end{equation}
Moreover, it is a measurement of angular current $J(\kappa)$, as 
the hydrodynamic equation Eq.~\ref{eq:J} at $\theta = -\pi$ gives
\begin{equation}
    \label{eq:current}
J(\kappa) = r_S\Omega_0|\varphi_S(-\pi)|^2 \left( \Delta/\delta -\kappa
\right).
\end{equation}

For $t > \tau_C$ radial rings will get superimposed on the spirals 
due to the self-interference, making extraction of curves of constant
$\xi(r,\theta)$ more difficult.  Moreover, when $t \sim \tau_S(r)$, the
derivatives of the initial angular wavefunction become important; finally,
for $t \gg \tau_S(r)$, the probability distribution resembles the Fourier
transform of the initial wavefunction, which has no spirals and the {\it in-situ} 
phase can not be read out.

\begin{figure*}
    \includegraphics[scale=1]{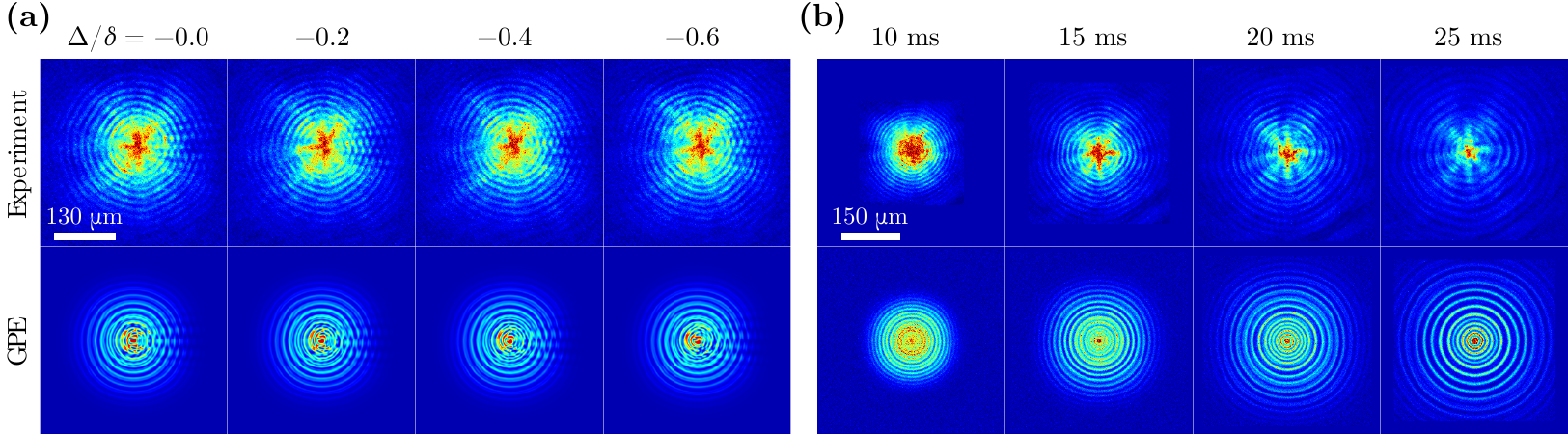} \caption{ 
    (Color online) Panel a) False color
    images of the interference pattern in the density profile after a
    $17$ ms expansion time for four rotation rates of the barrier.  
    The atom number
    density increases from blue to red with blue corresponding to zero
    density. 
    Top and bottom rows show images from our experiment and GPE
    simulations with the same trapping potentials and atom number, respectively. 
    The extracted $\Delta/\delta$ for each rotation rate is shown above the images.
    The winding number is zero for all images.
    Panel b) Images of experimental (top) and GPE (bottom)
    density profiles for four expansion times of a
    non-rotating condensate released from a target trap without a barrier.
}

\label{fig:exp_comp} \end{figure*}

\section{Experimental atom SQUID and mean-field simulation}
\label{sec:int}

We have also performed interference experiments with quantum-degenerate 
Sodium atoms in a target trap as well as simulations based on the
mean-field Gross-Pitaevskii equation (GPE). These results can be compared 
to our single-particle analysis and show the role of atom-atom
interactions present in ultra-cold atomtronic experiments.

The experimental setup is described in Sec.~\ref{subsec:experiment}.
Details of our numerical methods to simulate the GPE are given in
Sec.~\ref{subsec:Num_GP}, while Sec.~\ref{subsec:MF} describes expansion
timescales based on a self-similar expansion of a BEC from a target trap
\cite{Castin_Dum}. Section \ref{sec:expt} compares our results and enables
us to verify the extraction technique used in Ref.~\cite{Eckel2014b}
for the phase-drop across the barrier in terms of a measurement of
$\Delta/\delta$.

\subsection{Experimental setup}
\label{subsec:experiment}

We have performed interference experiments in a target trap.
We create a $^{23}\rm{Na}$ BEC in a target trap with approximately
$7\times 10^{5}$ atoms and a chemical potential, $\mu/\hbar \approx
2\pi\times(2\mbox{ kHz})$.  Details of the creation of the trapping
potential can be found in Refs.~\cite{Eckel2014, Eckel2014b}. The target
trap has an external toroid with a radius of 22.4(4) $\mu$m and radial
trapping frequency of 240 Hz. Its central disc has a flat-bottomed
potential and contains about 25\% of the total atoms. The transverse
trapping frequency of both traps is $\approx$ 600 Hz.  This leads to
a Bose condensate with a measured Thomas-Fermi radial width of about
6 $\mu$m in the toroid and a Thomas-Fermi radius of about 5 $\mu$m in
the disc.  The barrier potential has a Gaussian profile with a height less
than the chemical potential of the atoms in the science trap. Its $1/e^2$
full width is $\approx$ 6 $\mu$m.  Persistent current states are created
by adiabatically ramping up of the height of the barrier with a fixed
rotation rate. The atom cloud is imaged along the transverse direction
by absorption imaging, which measures the intensity of resonant light
transmitted through the expanding gas.

\subsection{Numerical simulation}
\label{subsec:Num_GP}

The initial wavefunction, $\Psi_{\rm GP}(\vec x)$, of the condensate
in the target trap is found in a two-step process. We first solve the
Gross-Pitaevskii equation for the wavefunction of a BEC with a {\it
stationary} weak link or barrier but otherwise the same trapping potentials and
atom number as in the experiment.  We use imaginary-time propagation
and a two-dimensional effective Lagrangian Variational Method (2D
LVM)~\cite{MarkLVM,OriginalLVM}, assuming a scattering length $a = 2.8$
nm.  The method is a variational technique whose trial wave function
is the product of an arbitrary function in the $(r, \theta)$ plane
and a Gaussian in the $z$ direction with an (imaginary-)time-dependent
width and a phase that is quadratic in $z$. This Ansatz leads to (a) a
two-dimensional effective GPE whose nonlinear coefficient contains the
width of the Gaussian and (b) an evolution equation for the width that
depends on the spatial integral of the fourth power of the absolute
value of the solution of the effective GPE.  We denote this solution
by $\Psi_{\rm Stat}(\vec x)$ and normalize such that $\int d^3\vec x
|\Psi_{\rm Stat}(\vec x)|^2=N$, the total atom number.  In particular,
we can find the angular density profile of the science trap $\rho_{\rm
Stat}(\theta)= \int' rdr dz|\Psi_{\rm Stat}(r,\theta,z)|^2$, where the
radial integral only encompasses the science or toroidal trap.

The second step is to add the rotation
of the barrier by multiplying the stationary (and
positive) $\Psi_{\rm Stat}(\vec x)$  with a spatially dependent phase
that leaves the density profile unchanged, i.e. $\Psi_{\rm GP}(\vec x) =
\Psi_{\rm Stat}(\vec x)e^{i \zeta(\vec x)}$. The phase profile $\zeta(\vec
x)$ is zero around and inside the central disk and near the torus only
depends on $\theta$. For a given rotation rate $\kappa$ and winding number $n$
it is found by simultaneously solving the hydrodynamic expression $J =
r_S\Omega_0 \rho_{\rm Stat}(\theta) (d\zeta(\theta)/d\theta - \kappa)$
and $\zeta(\pi) - \zeta(-\pi) = 2n\pi$.  (Compare to Eq.~\ref{eq:J} as
well as see the supplemental material in Ref.~\cite{Eckel2014b}). The
solution is similar in behavior to those shown in Fig.~\ref{fig:phase} and the
phase drop follows from $\gamma=2\pi(n-s)$, where $s=d\zeta(\theta)/d\theta\mid_{\theta=-\pi}$.

This phase-imprinting procedure is valid as long as the height of the
barrier is less than the chemical potential, the healing length $\ell =
\sqrt{\hbar/(2m\mu)}\approx 0.5\, \mu \rm m$ is small compared to the width of the barrier
($\approx 6$~$\mu$m),
and the speed of the barrier is small compared to the speed of sound $c =
\sqrt{\mu/m}$. These conditions are also met in the experiment.

Finally, we simulate the expansion of our BEC wavefunction released
from a target-trap by solving the (real) time-dependent Gross-Pitaevskii
equation using the same 2D-LVM method.  The GPE solutions have only been
modified to include the effects of absorption imaging.  The non-zero
point-spread-function of the imaging system is taken into account
by convolving the simulated transmission with an Airy disk of the
appropriate size.

\subsection{Expansion time scales}
\label{subsec:MF}

References \cite{Castin_Dum,kagan_evolution_1996} 
showed that a harmonically trapped and interacting Bose condensate
expands at a much faster rate than an non-interacting gas of the same
size. Here, we perform a similar analysis for expansion from a target
trap.  In fact, under the assumptions valid for phase imprinting in
Sec.~\ref{subsec:Num_GP}, it is sufficient to study expansion from a BEC
in a toroidal trap without a barrier or rotation.  We assume that the
interactions are sufficiently strong that the Thomas-Fermi approximation
holds along the $r$ and $z$ directions.  The BEC wavefunction is
then independent of $\theta$ and the harmonic confinement in the toroidal
trap along the $r$ and $z$ directions leads to a BEC with Thomas-Fermi
radius, $\sigma_{\rm TF}$, such that $\sigma_{\rm TF} \ll r_S$.  Here,
for simplicity we assume the same trap frequency along the two directions,
i.e.  $\omega_r = \omega_z \equiv \omega$.

Immediately, after the release of the toroidal trap the BEC expands
rapidly in the $r$ and $z$ directions as the interaction energy gets
converted to kinetic energy. This defines a ballistic timescale
$\tilde \tau_B$ (We use tilde to denote timescales associated
with expansion of the interacting BEC.)  As $\sigma_{\rm TF}
\ll r_S$, we can locally approximate an angular section of the
torus as a two dimensional tube, which expands along its transverse
directions.  Such an elongated BEC undergoes a self-similar expansion
\cite{Castin_Dum,kagan_evolution_1996}. That is, in the hydrodynamic
picture of the BEC and cylindrical coordinates, the density is $n(r,
z, t) \approx n(r_S + (r - r_S)/\lambda(t), z/\lambda(t), t=0)$ while
the velocity field $ \vec v(\vec x, t) = (v_r(r, t), 0, v_z(z, t))$
with $v_r(r, t) = (1 - \lambda(t)^{-2})(r-r_S)/t$  and $v_z(z, t) = (1 -
\lambda(t)^{-2})z/t$.  The scaling factor $\lambda(t) = \sqrt{1 + \omega^2
t^2}$, which  implies $\tilde\tau_B = 1/\omega = m \sigma_S^2/\hbar$ and
is the same as the single-particle ballistic time $\tau_B$, even though
the radial size of the BEC wavefunction $\sigma_{\rm TF} \gg \sigma_S$.

For $t \gg \tilde \tau_B$, the interaction energy has been converted
to kinetic energy, the density profile has spirals, but the cloud
is expanding more rapidly than the single-particle case.  Hence, we
expect that the time scale, $\tilde \tau_C(r)$, where the spirals become
modulated with circles due to the self-interference of the toroidal BEC, will
be shorter than the equivalent single-particle time scale, $\tau_C(r)$.
We can derive $\tilde\tau_C$ following the intuitive understanding of
signals from antipodal points $(r_S, \theta)$ and $(r_S, \theta + \pi)$
at $t=0$ reaching $(r, \theta)$ at $t=\tilde\tau_C$.  In other words,
we require that the radial size of the toroidal BEC, $\lambda(\tilde\tau_C)
\sigma_{\rm TF}$, is larger or equal to the distance between the
observation point and the antipodal points, i.e. $r+r_S$ and $r-r_S$.  Hence,
$\tilde \tau_C \approx  (r+r_S)/(\omega\sigma_{\rm TF})=(\sigma_S/\sigma_{\rm
TF}) \tau_C$, which is smaller than $\tau_C$.

\section{Comparison of the experiment with theory}
\label{sec:expt}
We compare our experimental data and GPE simulations in
Fig.~\ref{fig:exp_comp} by showing the dependence of the interference
pattern on the rotation rate of the barrier and the expansion time.
Figure \ref{fig:exp_comp}a) shows typical expanded clouds at $17$
ms expansion time from our experiment and simulated GPE expansions
for various rotation rates of the barrier leading to condensates with
winding number $n=0$. Firstly, we see radial interference fringes at
fixed $\theta$ and azimuthal interference fringes at fixed $r$ similar
to those in Fig.~\ref{fig:spirals}. The ratio $\Delta/\delta$ from these
experimental images is extracted following the procedure explained in
Fig.~\ref{fig:spirals}. The phase-drop across and the current through the
barrier then follows from Eqs.~\ref{eq:phase_drop} and \ref{eq:current},
respectively. Near $\theta =0$, where the barrier is located before
release, the density profile has radial stripes, which are absent from
the single-particle simulations and a consequence of interaction-induced
expansion of atoms into the density depleted weak-link region. Lastly,
star-like structures, which are due to residual azimuthal asymmetries
in the toroidal potential, are visible.

Figure \ref{fig:exp_comp}b) shows expanding, rotationless clouds released
from a trap without barrier for various expansion times.  For  observation
radii $r \geq 60$ $\mu$m and small expansion times $t \lesssim 20$ ms, the
experimental data and GPE results show no evidence of self-interference
of the toroidal BEC consistent with $t\le\tilde\tau_C(r)$.  For longer
expansion times we observe self-interference. It is prominent near the
cloud center, where radial fringes emerge with half the spacing of those
at large radius.

\begin{figure}
    \includegraphics[width=\columnwidth]{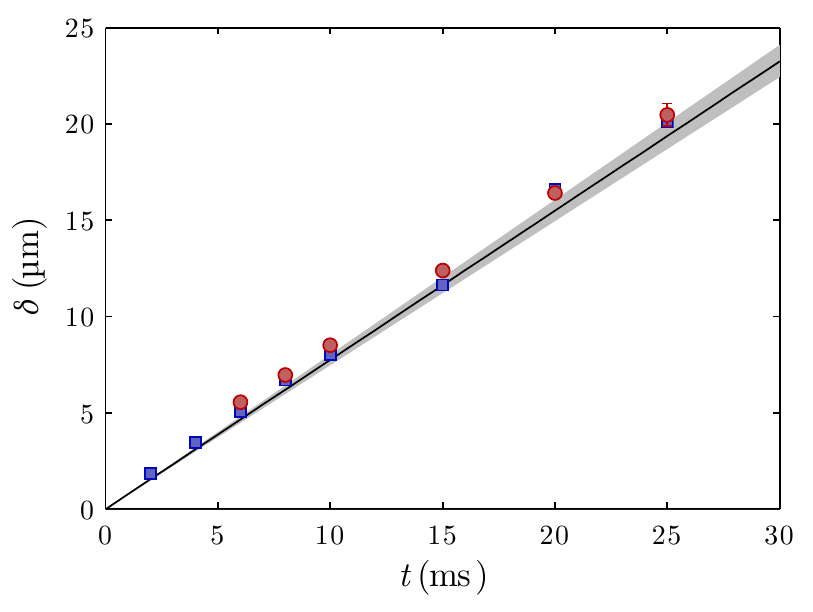} \caption{ Radial
    fringe spacing, $\delta$, of the interference pattern as a function
    of time elapsed after the release of the target trap.  The data is
    for a toroidal trap without a barrier and a BEC without winding.
    The experimental, GPE and single-particle fringe spacings are shown
    by red dots with one-standard deviation statistical error bars, blue
    markers and a black line, respectively. The value of
    $r_S$ has a uncertainty, which is shown by the shaded region around the black
    line.}
\label{fig:fringes} \end{figure}

In Fig.~\ref{fig:exp_comp} the size, shape and
interference pattern of the clouds in the GPE simulations agree well
with those of the experiment.  The agreement is made quantitative
in Fig.~\ref{fig:fringes} for the target trap without a barrier and
a BEC without winding ($n=0$). The figure shows the radial fringe
spacing, $\delta$, from the experimental data, GPE simulations and the
single-particle expression $\delta = 2\pi \hbar t/(m r_S)$ as functions
of expansion time. The three cases are in excellent agreement, indicating
that this fringe spacing is determined by the geometry of the system,
i.e. the radius of the torus.

\begin{figure}
	\center
    \includegraphics[width=\columnwidth]{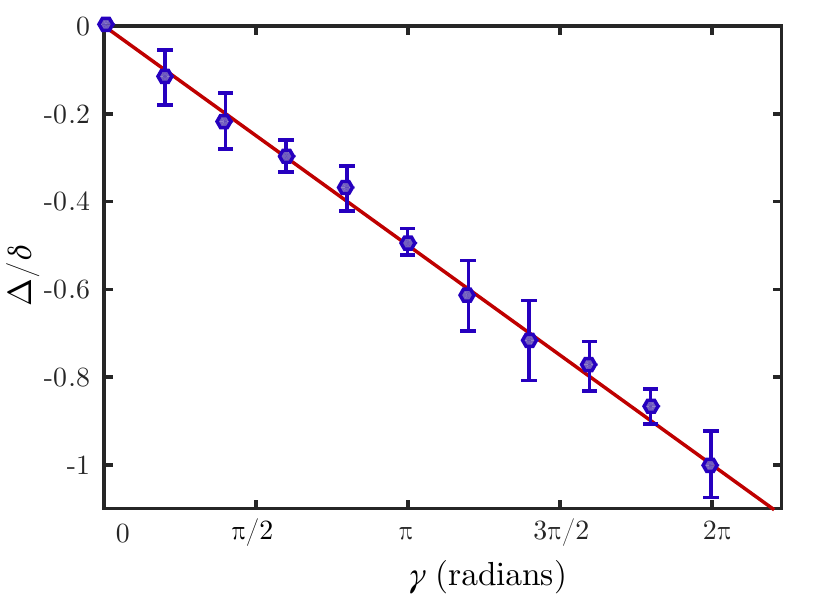}
	\caption{
    The ratio $\Delta/\delta$ as a function of {\it in-situ} phase-drop
    $\gamma$ across the rotating barrier from the GPE simulations of
    Fig.~\ref{fig:exp_comp}a) (markers) and the single-particle prediction (solid
    line). Error bars are one standard deviations uncertainties from
    the fit to the density profile.}
    \label{fig:encoded}
\end{figure}

Figure \ref{fig:encoded} shows the extracted $\Delta/\delta$ as a
function of the imprinted phase drop $\gamma$ across the barrier for the
GPE simulations in Fig.~\ref{fig:exp_comp}a).  The result agrees within
our uncertainties with the single-particle prediction, which indicates
that interactions do not change the phase drop over the barrier region
even though the angular density profile is distorted during the expansion.
In other words, an extraction of the phase drop from a measurement of
$\Delta/\delta$ is valid even when the GPE and experiment have radial
stripes for small $\theta$ near the weak link. The latter are absent
from the single-particle interference pattern.

\section{Conclusion}

We have experimentally and theoretically investigated an interferometric measurement
of the phase drop in an atomic-SQUID. The atomic-SQUID consists of a BEC in a toroidal
trap with a rotating barrier. The phase drop across the barrier is measured by
interference with a reference disc BEC after release from the trapping
potentials. We have studied the single-particle case and find that the
structure of the interference pattern depends on the expansion time after release.
For short times, it consists of spirals, which have the same number of
arms as the winding number of the toroidal wavefunction. The phase along a
spiral is the same as  the {\it in-situ} phase of the angular wavefunction.
Moreover, we find that the phase drop across the barrier and the current
through it determine the geometry of spirals.  For longer times
the spirals get superimposed by circles making phase readout difficult.

The conclusions from the single-particle model are confirmed by
experiments with Bose condensed sodium atoms and numerical simulations
based on the Gross-Pitaevskii equation even though inter-atomic
interactions speed up the expansion, thereby shortening the associated
time scales. In particular, one feature that is not changed is the fringe
spacings of the interference pattern.

Most importantly, we have confirmed that the phase-drop across the barrier
as measured by our experiment agree with those of our single-particle
model and mean-field simulations and accurately reflect the {\it in-situ} value.
This confirmation opens up the possibility of using this technique for
measuring the current-phase relationship of, for example, excitations or weak
links in degenerate, superfluid Fermi gases.

\bibliography{bib_rings}

\begin{thebibliography}{32}%
\makeatletter
\providecommand \@ifxundefined [1]{%
 \@ifx{#1\undefined}
}%
\providecommand \@ifnum [1]{%
 \ifnum #1\expandafter \@firstoftwo
 \else \expandafter \@secondoftwo
 \fi
}%
\providecommand \@ifx [1]{%
 \ifx #1\expandafter \@firstoftwo
 \else \expandafter \@secondoftwo
 \fi
}%
\providecommand \natexlab [1]{#1}%
\providecommand \enquote  [1]{``#1''}%
\providecommand \bibnamefont  [1]{#1}%
\providecommand \bibfnamefont [1]{#1}%
\providecommand \citenamefont [1]{#1}%
\providecommand \href@noop [0]{\@secondoftwo}%
\providecommand \href [0]{\begingroup \@sanitize@url \@href}%
\providecommand \@href[1]{\@@startlink{#1}\@@href}%
\providecommand \@@href[1]{\endgroup#1\@@endlink}%
\providecommand \@sanitize@url [0]{\catcode `\\12\catcode `\$12\catcode
  `\&12\catcode `\#12\catcode `\^12\catcode `\_12\catcode `\%12\relax}%
\providecommand \@@startlink[1]{}%
\providecommand \@@endlink[0]{}%
\providecommand \url  [0]{\begingroup\@sanitize@url \@url }%
\providecommand \@url [1]{\endgroup\@href {#1}{\urlprefix }}%
\providecommand \urlprefix  [0]{URL }%
\providecommand \Eprint [0]{\href }%
\providecommand \doibase [0]{http://dx.doi.org/}%
\providecommand \selectlanguage [0]{\@gobble}%
\providecommand \bibinfo  [0]{\@secondoftwo}%
\providecommand \bibfield  [0]{\@secondoftwo}%
\providecommand \translation [1]{[#1]}%
\providecommand \BibitemOpen [0]{}%
\providecommand \bibitemStop [0]{}%
\providecommand \bibitemNoStop [0]{.\EOS\space}%
\providecommand \EOS [0]{\spacefactor3000\relax}%
\providecommand \BibitemShut  [1]{\csname bibitem#1\endcsname}%
\let\auto@bib@innerbib\@empty
\bibitem [{\citenamefont {Seaman}\ \emph {et~al.}(2007)\citenamefont {Seaman},
  \citenamefont {Kr\"amer}, \citenamefont {Anderson},\ and\ \citenamefont
  {Holland}}]{seaman_2007}%
  \BibitemOpen
  \bibfield  {author} {\bibinfo {author} {\bibfnamefont {B.~T.}\ \bibnamefont
  {Seaman}}, \bibinfo {author} {\bibfnamefont {M.}~\bibnamefont {Kr\"amer}},
  \bibinfo {author} {\bibfnamefont {D.~Z.}\ \bibnamefont {Anderson}}, \ and\
  \bibinfo {author} {\bibfnamefont {M.~J.}\ \bibnamefont {Holland}},\ }\href
  {http://link.aps.org/doi/10.1103/PhysRevA.75.023615} {\bibfield  {journal}
  {\bibinfo  {journal} {Phys. Rev. A}\ }\textbf {\bibinfo {volume} {75}},\
  \bibinfo {pages} {023615} (\bibinfo {year} {2007})}\BibitemShut {NoStop}%
\bibitem [{\citenamefont {Lee}\ \emph {et~al.}(2013)\citenamefont {Lee},
  \citenamefont {McIlvain}, \citenamefont {Lobb},\ and\ \citenamefont
  {Hill}}]{Lee2013}%
  \BibitemOpen
  \bibfield  {author} {\bibinfo {author} {\bibfnamefont {J.~G.}\ \bibnamefont
  {Lee}}, \bibinfo {author} {\bibfnamefont {B.~J.}\ \bibnamefont {McIlvain}},
  \bibinfo {author} {\bibfnamefont {C.~J.}\ \bibnamefont {Lobb}}, \ and\
  \bibinfo {author} {\bibfnamefont {I.}~\bibnamefont {Hill}, \bibfnamefont
  {W.~T.}},\ }\href {http://dx.doi.org/10.1038/srep01034} {\bibfield  {journal}
  {\bibinfo  {journal} {Sci. Rep.}\ }\textbf {\bibinfo {volume} {3}},\ \bibinfo
  {pages} {1034} (\bibinfo {year} {2013})}\BibitemShut {NoStop}%
\bibitem [{\citenamefont {Krinner}\ \emph {et~al.}(2015)\citenamefont
  {Krinner}, \citenamefont {Stadler}, \citenamefont {Husmann}, \citenamefont
  {Brantut},\ and\ \citenamefont {Esslinger}}]{krinner_2015}%
  \BibitemOpen
  \bibfield  {author} {\bibinfo {author} {\bibfnamefont {S.}~\bibnamefont
  {Krinner}}, \bibinfo {author} {\bibfnamefont {D.}~\bibnamefont {Stadler}},
  \bibinfo {author} {\bibfnamefont {D.}~\bibnamefont {Husmann}}, \bibinfo
  {author} {\bibfnamefont {J.-P.}\ \bibnamefont {Brantut}}, \ and\ \bibinfo
  {author} {\bibfnamefont {T.}~\bibnamefont {Esslinger}},\ }\href {\doibase
  10.1038/nature14049} {\bibfield  {journal} {\bibinfo  {journal} {Nature}\
  }\textbf {\bibinfo {volume} {517}},\ \bibinfo {pages} {64} (\bibinfo {year}
  {2015})}\BibitemShut {NoStop}%
\bibitem [{\citenamefont {Beeler}\ \emph {et~al.}(2013)\citenamefont {Beeler},
  \citenamefont {Williams}, \citenamefont {Jim\'enez-Garc\'ia}, \citenamefont
  {LeBlanc}, \citenamefont {Perry},\ and\ \citenamefont
  {Spielman}}]{Beeler2013}%
  \BibitemOpen
  \bibfield  {author} {\bibinfo {author} {\bibfnamefont {M.~C.}\ \bibnamefont
  {Beeler}}, \bibinfo {author} {\bibfnamefont {R.~A.}\ \bibnamefont
  {Williams}}, \bibinfo {author} {\bibfnamefont {K.}~\bibnamefont
  {Jim\'enez-Garc\'ia}}, \bibinfo {author} {\bibfnamefont {L.~J.}\ \bibnamefont
  {LeBlanc}}, \bibinfo {author} {\bibfnamefont {A.~R.}\ \bibnamefont {Perry}},
  \ and\ \bibinfo {author} {\bibfnamefont {I.~B.}\ \bibnamefont {Spielman}},\
  }\href {\doibase 10.1038/nature12185} {\bibfield  {journal} {\bibinfo
  {journal} {Nature}\ }\textbf {\bibinfo {volume} {498}},\ \bibinfo {pages}
  {201} (\bibinfo {year} {2013})}\BibitemShut {NoStop}%
\bibitem [{\citenamefont {Ramanathan}\ \emph {et~al.}(2011)\citenamefont
  {Ramanathan}, \citenamefont {Wright}, \citenamefont {Muniz}, \citenamefont
  {Zelan}, \citenamefont {Hill}, \citenamefont {Lobb}, \citenamefont
  {Helmerson}, \citenamefont {Phillips},\ and\ \citenamefont
  {Campbell}}]{Ramanathan2011}%
  \BibitemOpen
  \bibfield  {author} {\bibinfo {author} {\bibfnamefont {A.}~\bibnamefont
  {Ramanathan}}, \bibinfo {author} {\bibfnamefont {K.~C.}\ \bibnamefont
  {Wright}}, \bibinfo {author} {\bibfnamefont {S.~R.}\ \bibnamefont {Muniz}},
  \bibinfo {author} {\bibfnamefont {M.}~\bibnamefont {Zelan}}, \bibinfo
  {author} {\bibfnamefont {W.~T.}\ \bibnamefont {Hill}}, \bibinfo {author}
  {\bibfnamefont {C.~J.}\ \bibnamefont {Lobb}}, \bibinfo {author}
  {\bibfnamefont {K.}~\bibnamefont {Helmerson}}, \bibinfo {author}
  {\bibfnamefont {W.~D.}\ \bibnamefont {Phillips}}, \ and\ \bibinfo {author}
  {\bibfnamefont {G.~K.}\ \bibnamefont {Campbell}},\ }\href {\doibase
  10.1103/PhysRevLett.106.130401} {\bibfield  {journal} {\bibinfo  {journal}
  {Phys. Rev. Lett.}\ }\textbf {\bibinfo {volume} {106}},\ \bibinfo {pages}
  {130401} (\bibinfo {year} {2011})}\BibitemShut {NoStop}%
\bibitem [{\citenamefont {Wright}\ \emph {et~al.}(2013)\citenamefont {Wright},
  \citenamefont {Blakestad}, \citenamefont {Lobb}, \citenamefont {Phillips},\
  and\ \citenamefont {Campbell}}]{wright_2013}%
  \BibitemOpen
  \bibfield  {author} {\bibinfo {author} {\bibfnamefont {K.~C.}\ \bibnamefont
  {Wright}}, \bibinfo {author} {\bibfnamefont {R.~B.}\ \bibnamefont
  {Blakestad}}, \bibinfo {author} {\bibfnamefont {C.~J.}\ \bibnamefont {Lobb}},
  \bibinfo {author} {\bibfnamefont {W.~D.}\ \bibnamefont {Phillips}}, \ and\
  \bibinfo {author} {\bibfnamefont {G.~K.}\ \bibnamefont {Campbell}},\ }\href
  {\doibase 10.1103/PhysRevLett.110.025302} {\bibfield  {journal} {\bibinfo
  {journal} {Phys. Rev. Lett.}\ }\textbf {\bibinfo {volume} {110}},\ \bibinfo
  {pages} {025302} (\bibinfo {year} {2013})}\BibitemShut {NoStop}%
\bibitem [{\citenamefont {Eckel}\ \emph
  {et~al.}(2014{\natexlab{a}})\citenamefont {Eckel}, \citenamefont {Lee},
  \citenamefont {Jendrzejewski}, \citenamefont {Murray}, \citenamefont {Clark},
  \citenamefont {Lobb}, \citenamefont {Phillips}, \citenamefont {Edwards},\
  and\ \citenamefont {Campbell}}]{Eckel2014}%
  \BibitemOpen
  \bibfield  {author} {\bibinfo {author} {\bibfnamefont {S.}~\bibnamefont
  {Eckel}}, \bibinfo {author} {\bibfnamefont {J.~G.}\ \bibnamefont {Lee}},
  \bibinfo {author} {\bibfnamefont {F.}~\bibnamefont {Jendrzejewski}}, \bibinfo
  {author} {\bibfnamefont {N.}~\bibnamefont {Murray}}, \bibinfo {author}
  {\bibfnamefont {C.~W.}\ \bibnamefont {Clark}}, \bibinfo {author}
  {\bibfnamefont {C.~J.}\ \bibnamefont {Lobb}}, \bibinfo {author}
  {\bibfnamefont {W.~D.}\ \bibnamefont {Phillips}}, \bibinfo {author}
  {\bibfnamefont {M.}~\bibnamefont {Edwards}}, \ and\ \bibinfo {author}
  {\bibfnamefont {G.~K.}\ \bibnamefont {Campbell}},\ }\href {\doibase
  10.1038/nature12958} {\bibfield  {journal} {\bibinfo  {journal} {Nature}\
  }\textbf {\bibinfo {volume} {506}},\ \bibinfo {pages} {200} (\bibinfo {year}
  {2014}{\natexlab{a}})}\BibitemShut {NoStop}%
\bibitem [{\citenamefont {Ryu}\ \emph {et~al.}(2013)\citenamefont {Ryu},
  \citenamefont {Blackburn}, \citenamefont {Blinova},\ and\ \citenamefont
  {Boshier}}]{Ryu2013}%
  \BibitemOpen
  \bibfield  {author} {\bibinfo {author} {\bibfnamefont {C.}~\bibnamefont
  {Ryu}}, \bibinfo {author} {\bibfnamefont {P.~W.}\ \bibnamefont {Blackburn}},
  \bibinfo {author} {\bibfnamefont {A.~A.}\ \bibnamefont {Blinova}}, \ and\
  \bibinfo {author} {\bibfnamefont {M.~G.}\ \bibnamefont {Boshier}},\ }\href
  {\doibase 10.1103/PhysRevLett.111.205301} {\bibfield  {journal} {\bibinfo
  {journal} {Phys. Rev. Lett.}\ }\textbf {\bibinfo {volume} {111}},\ \bibinfo
  {pages} {205301} (\bibinfo {year} {2013})}\BibitemShut {NoStop}%
\bibitem [{\citenamefont {Jendrzejewski}\ \emph {et~al.}(2014)\citenamefont
  {Jendrzejewski}, \citenamefont {Eckel}, \citenamefont {Murray}, \citenamefont
  {Lanier}, \citenamefont {Edwards}, \citenamefont {Lobb},\ and\ \citenamefont
  {Campbell}}]{Jendrzejewski2014}%
  \BibitemOpen
  \bibfield  {author} {\bibinfo {author} {\bibfnamefont {F.}~\bibnamefont
  {Jendrzejewski}}, \bibinfo {author} {\bibfnamefont {S.}~\bibnamefont
  {Eckel}}, \bibinfo {author} {\bibfnamefont {N.}~\bibnamefont {Murray}},
  \bibinfo {author} {\bibfnamefont {C.}~\bibnamefont {Lanier}}, \bibinfo
  {author} {\bibfnamefont {M.}~\bibnamefont {Edwards}}, \bibinfo {author}
  {\bibfnamefont {C.~J.}\ \bibnamefont {Lobb}}, \ and\ \bibinfo {author}
  {\bibfnamefont {G.~K.}\ \bibnamefont {Campbell}},\ }\href {\doibase
  10.1103/PhysRevLett.113.045305} {\bibfield  {journal} {\bibinfo  {journal}
  {Phys. Rev. Lett.}\ }\textbf {\bibinfo {volume} {113}},\ \bibinfo {pages}
  {045305} (\bibinfo {year} {2014})}\BibitemShut {NoStop}%
\bibitem [{\citenamefont {Amico}\ \emph {et~al.}(2014)\citenamefont {Amico},
  \citenamefont {Aghamalyan}, \citenamefont {Auksztol}, \citenamefont {Crepaz},
  \citenamefont {Dumke},\ and\ \citenamefont {Kwek}}]{amico_superfluid_2014}%
  \BibitemOpen
  \bibfield  {author} {\bibinfo {author} {\bibfnamefont {L.}~\bibnamefont
  {Amico}}, \bibinfo {author} {\bibfnamefont {D.}~\bibnamefont {Aghamalyan}},
  \bibinfo {author} {\bibfnamefont {F.}~\bibnamefont {Auksztol}}, \bibinfo
  {author} {\bibfnamefont {H.}~\bibnamefont {Crepaz}}, \bibinfo {author}
  {\bibfnamefont {R.}~\bibnamefont {Dumke}}, \ and\ \bibinfo {author}
  {\bibfnamefont {L.~C.}\ \bibnamefont {Kwek}},\ }\href {\doibase
  10.1038/srep04298} {\bibfield  {journal} {\bibinfo  {journal} {Sci. Rep.}\
  }\textbf {\bibinfo {volume} {4}},\ \bibinfo {pages} {4298} (\bibinfo {year}
  {2014})}\BibitemShut {NoStop}%
\bibitem [{\citenamefont {Aghamalyan}\ \emph {et~al.}(2015)\citenamefont
  {Aghamalyan}, \citenamefont {Cominotti}, \citenamefont {Rizzi}, \citenamefont
  {Rossini}, \citenamefont {Hekking}, \citenamefont {Minguzzi}, \citenamefont
  {Kwek},\ and\ \citenamefont {Amico}}]{aghamalyan_coherent_2015}%
  \BibitemOpen
  \bibfield  {author} {\bibinfo {author} {\bibfnamefont {D.}~\bibnamefont
  {Aghamalyan}}, \bibinfo {author} {\bibfnamefont {M.}~\bibnamefont
  {Cominotti}}, \bibinfo {author} {\bibfnamefont {M.}~\bibnamefont {Rizzi}},
  \bibinfo {author} {\bibfnamefont {D.}~\bibnamefont {Rossini}}, \bibinfo
  {author} {\bibfnamefont {F.}~\bibnamefont {Hekking}}, \bibinfo {author}
  {\bibfnamefont {A.}~\bibnamefont {Minguzzi}}, \bibinfo {author}
  {\bibfnamefont {L.-C.}\ \bibnamefont {Kwek}}, \ and\ \bibinfo {author}
  {\bibfnamefont {L.}~\bibnamefont {Amico}},\ }\href {\doibase
  10.1088/1367-2630/17/4/045023} {\bibfield  {journal} {\bibinfo  {journal}
  {New J. Phys.}\ }\textbf {\bibinfo {volume} {17}},\ \bibinfo {pages} {045023}
  (\bibinfo {year} {2015})}\BibitemShut {NoStop}%
\bibitem [{\citenamefont {Kashurnikov}\ \emph {et~al.}(1996)\citenamefont
  {Kashurnikov}, \citenamefont {Podlivaev}, \citenamefont {Prokof'ev},\ and\
  \citenamefont {Svistunov}}]{kashurnikov_1996}%
  \BibitemOpen
  \bibfield  {author} {\bibinfo {author} {\bibfnamefont {V.~A.}\ \bibnamefont
  {Kashurnikov}}, \bibinfo {author} {\bibfnamefont {A.~I.}\ \bibnamefont
  {Podlivaev}}, \bibinfo {author} {\bibfnamefont {N.~V.}\ \bibnamefont
  {Prokof'ev}}, \ and\ \bibinfo {author} {\bibfnamefont {B.~V.}\ \bibnamefont
  {Svistunov}},\ }\href {\doibase 10.1103/PhysRevB.53.13091} {\bibfield
  {journal} {\bibinfo  {journal} {Phys. Rev. B}\ }\textbf {\bibinfo {volume}
  {53}},\ \bibinfo {pages} {13091} (\bibinfo {year} {1996})}\BibitemShut
  {NoStop}%
\bibitem [{\citenamefont {Kagan}\ \emph {et~al.}(2000)\citenamefont {Kagan},
  \citenamefont {Prokof’ev},\ and\ \citenamefont {Svistunov}}]{kagan_2000}%
  \BibitemOpen
  \bibfield  {author} {\bibinfo {author} {\bibfnamefont {Y.}~\bibnamefont
  {Kagan}}, \bibinfo {author} {\bibfnamefont {N.~V.}\ \bibnamefont
  {Prokof’ev}}, \ and\ \bibinfo {author} {\bibfnamefont {B.~V.}\ \bibnamefont
  {Svistunov}},\ }\href {\doibase 10.1103/PhysRevA.61.045601} {\bibfield
  {journal} {\bibinfo  {journal} {Phys. Rev. A}\ }\textbf {\bibinfo {volume}
  {61}},\ \bibinfo {pages} {045601} (\bibinfo {year} {2000})}\BibitemShut
  {NoStop}%
\bibitem [{\citenamefont {B\"uchler}\ \emph {et~al.}(2001)\citenamefont
  {B\"uchler}, \citenamefont {Geshkenbein},\ and\ \citenamefont
  {Blatter}}]{buchler_2001}%
  \BibitemOpen
  \bibfield  {author} {\bibinfo {author} {\bibfnamefont {H.~P.}\ \bibnamefont
  {B\"uchler}}, \bibinfo {author} {\bibfnamefont {V.~B.}\ \bibnamefont
  {Geshkenbein}}, \ and\ \bibinfo {author} {\bibfnamefont {G.}~\bibnamefont
  {Blatter}},\ }\href {\doibase 10.1103/PhysRevLett.87.100403} {\bibfield
  {journal} {\bibinfo  {journal} {Phys. Rev. Lett.}\ }\textbf {\bibinfo
  {volume} {87}},\ \bibinfo {pages} {100403} (\bibinfo {year}
  {2001})}\BibitemShut {NoStop}%
\bibitem [{\citenamefont {Cominotti}\ \emph {et~al.}(2014)\citenamefont
  {Cominotti}, \citenamefont {Rossini}, \citenamefont {Rizzi}, \citenamefont
  {Hekking},\ and\ \citenamefont {Minguzzi}}]{cominotti_2014}%
  \BibitemOpen
  \bibfield  {author} {\bibinfo {author} {\bibfnamefont {M.}~\bibnamefont
  {Cominotti}}, \bibinfo {author} {\bibfnamefont {D.}~\bibnamefont {Rossini}},
  \bibinfo {author} {\bibfnamefont {M.}~\bibnamefont {Rizzi}}, \bibinfo
  {author} {\bibfnamefont {F.}~\bibnamefont {Hekking}}, \ and\ \bibinfo
  {author} {\bibfnamefont {A.}~\bibnamefont {Minguzzi}},\ }\href {\doibase
  10.1103/PhysRevLett.113.025301} {\bibfield  {journal} {\bibinfo  {journal}
  {Phys. Rev. Lett.}\ }\textbf {\bibinfo {volume} {113}},\ \bibinfo {pages}
  {025301} (\bibinfo {year} {2014})}\BibitemShut {NoStop}%
\bibitem [{\citenamefont {Likharev}(1979)}]{Likharex1979}%
  \BibitemOpen
  \bibfield  {author} {\bibinfo {author} {\bibfnamefont {K.}~\bibnamefont
  {Likharev}},\ }\href {\doibase 10.1103/RevModPhys.51.101} {\bibfield
  {journal} {\bibinfo  {journal} {Rev. Mod. Phys.}\ }\textbf {\bibinfo {volume}
  {51}},\ \bibinfo {pages} {101} (\bibinfo {year} {1979})}\BibitemShut
  {NoStop}%
\bibitem [{\citenamefont {Andrews}\ \emph {et~al.}(1997)\citenamefont
  {Andrews}, \citenamefont {Townsend}, \citenamefont {Miesner}, \citenamefont
  {Durfree}, \citenamefont {Kurn},\ and\ \citenamefont
  {Ketterle}}]{Andrews1997}%
  \BibitemOpen
  \bibfield  {author} {\bibinfo {author} {\bibfnamefont {M.~R.}\ \bibnamefont
  {Andrews}}, \bibinfo {author} {\bibfnamefont {C.~G.}\ \bibnamefont
  {Townsend}}, \bibinfo {author} {\bibfnamefont {H.-J.}\ \bibnamefont
  {Miesner}}, \bibinfo {author} {\bibfnamefont {D.~S.}\ \bibnamefont
  {Durfree}}, \bibinfo {author} {\bibfnamefont {D.~M.}\ \bibnamefont {Kurn}}, \
  and\ \bibinfo {author} {\bibfnamefont {W.}~\bibnamefont {Ketterle}},\ }\href
  {\doibase 10.1126/science.275.5300.637} {\bibfield  {journal} {\bibinfo
  {journal} {Science}\ }\textbf {\bibinfo {volume} {275}},\ \bibinfo {pages}
  {637} (\bibinfo {year} {1997})}\BibitemShut {NoStop}%
\bibitem [{\citenamefont {Castin}\ and\ \citenamefont
  {Dalibard}(1997)}]{Castin1997}%
  \BibitemOpen
  \bibfield  {author} {\bibinfo {author} {\bibfnamefont {Y.}~\bibnamefont
  {Castin}}\ and\ \bibinfo {author} {\bibfnamefont {J.}~\bibnamefont
  {Dalibard}},\ }\href {\doibase 10.1103/PhysRevA.55.4330} {\bibfield
  {journal} {\bibinfo  {journal} {Phys. Rev. A}\ }\textbf {\bibinfo {volume}
  {55}},\ \bibinfo {pages} {4330} (\bibinfo {year} {1997})}\BibitemShut
  {NoStop}%
\bibitem [{\citenamefont {Simsarian}\ \emph {et~al.}(2000)\citenamefont
  {Simsarian}, \citenamefont {Denschlag}, \citenamefont {Edwards},
  \citenamefont {Clark}, \citenamefont {Deng}, \citenamefont {Hagley},
  \citenamefont {Helmerson}, \citenamefont {Rolston},\ and\ \citenamefont
  {Phillips}}]{simsarian_imaging_2000}%
  \BibitemOpen
  \bibfield  {author} {\bibinfo {author} {\bibfnamefont {J.~E.}\ \bibnamefont
  {Simsarian}}, \bibinfo {author} {\bibfnamefont {J.}~\bibnamefont
  {Denschlag}}, \bibinfo {author} {\bibfnamefont {M.}~\bibnamefont {Edwards}},
  \bibinfo {author} {\bibfnamefont {C.~W.}\ \bibnamefont {Clark}}, \bibinfo
  {author} {\bibfnamefont {L.}~\bibnamefont {Deng}}, \bibinfo {author}
  {\bibfnamefont {E.~W.}\ \bibnamefont {Hagley}}, \bibinfo {author}
  {\bibfnamefont {K.}~\bibnamefont {Helmerson}}, \bibinfo {author}
  {\bibfnamefont {S.~L.}\ \bibnamefont {Rolston}}, \ and\ \bibinfo {author}
  {\bibfnamefont {W.~D.}\ \bibnamefont {Phillips}},\ }\href {\doibase
  10.1103/PhysRevLett.85.2040} {\bibfield  {journal} {\bibinfo  {journal}
  {Phys. Rev. Lett.}\ }\textbf {\bibinfo {volume} {85}},\ \bibinfo {pages}
  {2040} (\bibinfo {year} {2000})}\BibitemShut {NoStop}%
\bibitem [{\citenamefont {Inouye}\ \emph {et~al.}(2001)\citenamefont {Inouye},
  \citenamefont {Gupta}, \citenamefont {Rosenband}, \citenamefont {Chikkatur},
  \citenamefont {G\"{o}rlitz}, \citenamefont {Gustavson}, \citenamefont
  {Leanhardt}, \citenamefont {Pritchard},\ and\ \citenamefont
  {Ketterle}}]{inouye_2001}%
  \BibitemOpen
  \bibfield  {author} {\bibinfo {author} {\bibfnamefont {S.}~\bibnamefont
  {Inouye}}, \bibinfo {author} {\bibfnamefont {S.}~\bibnamefont {Gupta}},
  \bibinfo {author} {\bibfnamefont {T.}~\bibnamefont {Rosenband}}, \bibinfo
  {author} {\bibfnamefont {A.~P.}\ \bibnamefont {Chikkatur}}, \bibinfo {author}
  {\bibfnamefont {A.}~\bibnamefont {G\"{o}rlitz}}, \bibinfo {author}
  {\bibfnamefont {T.~L.}\ \bibnamefont {Gustavson}}, \bibinfo {author}
  {\bibfnamefont {A.~E.}\ \bibnamefont {Leanhardt}}, \bibinfo {author}
  {\bibfnamefont {D.~E.}\ \bibnamefont {Pritchard}}, \ and\ \bibinfo {author}
  {\bibfnamefont {W.}~\bibnamefont {Ketterle}},\ }\href {\doibase
  10.1103/PhysRevLett.87.080402} {\bibfield  {journal} {\bibinfo  {journal}
  {Phys. Rev. Lett.}\ }\textbf {\bibinfo {volume} {87}},\ \bibinfo {pages}
  {080402} (\bibinfo {year} {2001})}\BibitemShut {NoStop}%
\bibitem [{\citenamefont {Hadzibabic}\ \emph {et~al.}(2006)\citenamefont
  {Hadzibabic}, \citenamefont {Kr\"{u}ger}, \citenamefont {Cheneau},
  \citenamefont {Battelier},\ and\ \citenamefont {Dalibard}}]{Hadzibabic2006}%
  \BibitemOpen
  \bibfield  {author} {\bibinfo {author} {\bibfnamefont {Z.}~\bibnamefont
  {Hadzibabic}}, \bibinfo {author} {\bibfnamefont {P.}~\bibnamefont
  {Kr\"{u}ger}}, \bibinfo {author} {\bibfnamefont {M.}~\bibnamefont {Cheneau}},
  \bibinfo {author} {\bibfnamefont {B.}~\bibnamefont {Battelier}}, \ and\
  \bibinfo {author} {\bibfnamefont {J.}~\bibnamefont {Dalibard}},\ }\href
  {\doibase 10.1038/nature04851} {\bibfield  {journal} {\bibinfo  {journal}
  {Nature}\ }\textbf {\bibinfo {volume} {441}},\ \bibinfo {pages} {1118}
  (\bibinfo {year} {2006})}\BibitemShut {NoStop}%
\bibitem [{\citenamefont {Kohstall}\ \emph {et~al.}(2011)\citenamefont
  {Kohstall}, \citenamefont {Riedl}, \citenamefont {{S\'{a}nchez Guajardo}},
  \citenamefont {Sidorenkov}, \citenamefont {{Hecker Denschlag}},\ and\
  \citenamefont {Grimm}}]{Kohstall2011}%
  \BibitemOpen
  \bibfield  {author} {\bibinfo {author} {\bibfnamefont {C.}~\bibnamefont
  {Kohstall}}, \bibinfo {author} {\bibfnamefont {S.}~\bibnamefont {Riedl}},
  \bibinfo {author} {\bibfnamefont {E.~R.}\ \bibnamefont {{S\'{a}nchez
  Guajardo}}}, \bibinfo {author} {\bibfnamefont {L.~A.}\ \bibnamefont
  {Sidorenkov}}, \bibinfo {author} {\bibfnamefont {J.}~\bibnamefont {{Hecker
  Denschlag}}}, \ and\ \bibinfo {author} {\bibfnamefont {R.}~\bibnamefont
  {Grimm}},\ }\href {\doibase 10.1088/1367-2630/13/6/065027} {\bibfield
  {journal} {\bibinfo  {journal} {New J. Phys.}\ }\textbf {\bibinfo {volume}
  {13}},\ \bibinfo {pages} {065027} (\bibinfo {year} {2011})}\BibitemShut
  {NoStop}%
\bibitem [{\citenamefont {Shin}\ \emph {et~al.}(2005)\citenamefont {Shin},
  \citenamefont {Sanner}, \citenamefont {Jo}, \citenamefont {Pasquini},
  \citenamefont {Saba}, \citenamefont {Ketterle}, \citenamefont {Pritchard},
  \citenamefont {Vengalattore},\ and\ \citenamefont {Prentiss}}]{Shin2005}%
  \BibitemOpen
  \bibfield  {author} {\bibinfo {author} {\bibfnamefont {Y.}~\bibnamefont
  {Shin}}, \bibinfo {author} {\bibfnamefont {C.}~\bibnamefont {Sanner}},
  \bibinfo {author} {\bibfnamefont {G.-B.}\ \bibnamefont {Jo}}, \bibinfo
  {author} {\bibfnamefont {T.~A.}\ \bibnamefont {Pasquini}}, \bibinfo {author}
  {\bibfnamefont {M.}~\bibnamefont {Saba}}, \bibinfo {author} {\bibfnamefont
  {W.}~\bibnamefont {Ketterle}}, \bibinfo {author} {\bibfnamefont {D.~E.}\
  \bibnamefont {Pritchard}}, \bibinfo {author} {\bibfnamefont {M.}~\bibnamefont
  {Vengalattore}}, \ and\ \bibinfo {author} {\bibfnamefont {M.}~\bibnamefont
  {Prentiss}},\ }\href {\doibase 10.1103/PhysRevA.72.021604} {\bibfield
  {journal} {\bibinfo  {journal} {Phys. Rev. A}\ }\textbf {\bibinfo {volume}
  {72}},\ \bibinfo {pages} {021604} (\bibinfo {year} {2005})}\BibitemShut
  {NoStop}%
\bibitem [{\citenamefont {Corman}\ \emph {et~al.}(2014)\citenamefont {Corman},
  \citenamefont {Chomaz}, \citenamefont {Bienaim\'e}, \citenamefont
  {Desbuquois}, \citenamefont {Weitenberg}, \citenamefont {Nascimb\`ene},
  \citenamefont {Dalibard},\ and\ \citenamefont {Beugnon}}]{Corman2014}%
  \BibitemOpen
  \bibfield  {author} {\bibinfo {author} {\bibfnamefont {L.}~\bibnamefont
  {Corman}}, \bibinfo {author} {\bibfnamefont {L.}~\bibnamefont {Chomaz}},
  \bibinfo {author} {\bibfnamefont {T.}~\bibnamefont {Bienaim\'e}}, \bibinfo
  {author} {\bibfnamefont {R.}~\bibnamefont {Desbuquois}}, \bibinfo {author}
  {\bibfnamefont {C.}~\bibnamefont {Weitenberg}}, \bibinfo {author}
  {\bibfnamefont {S.}~\bibnamefont {Nascimb\`ene}}, \bibinfo {author}
  {\bibfnamefont {J.}~\bibnamefont {Dalibard}}, \ and\ \bibinfo {author}
  {\bibfnamefont {J.}~\bibnamefont {Beugnon}},\ }\href {\doibase
  10.1103/PhysRevLett.113.135302} {\bibfield  {journal} {\bibinfo  {journal}
  {Phys. Rev. Lett.}\ }\textbf {\bibinfo {volume} {113}},\ \bibinfo {pages}
  {135302} (\bibinfo {year} {2014})}\BibitemShut {NoStop}%
\bibitem [{\citenamefont {Eckel}\ \emph
  {et~al.}(2014{\natexlab{b}})\citenamefont {Eckel}, \citenamefont
  {Jendrzejewski}, \citenamefont {Kumar}, \citenamefont {Lobb},\ and\
  \citenamefont {Campbell}}]{Eckel2014b}%
  \BibitemOpen
  \bibfield  {author} {\bibinfo {author} {\bibfnamefont {S.}~\bibnamefont
  {Eckel}}, \bibinfo {author} {\bibfnamefont {F.}~\bibnamefont
  {Jendrzejewski}}, \bibinfo {author} {\bibfnamefont {A.}~\bibnamefont
  {Kumar}}, \bibinfo {author} {\bibfnamefont {C.}~\bibnamefont {Lobb}}, \ and\
  \bibinfo {author} {\bibfnamefont {G.}~\bibnamefont {Campbell}},\ }\href
  {\doibase 10.1103/PhysRevX.4.031052} {\bibfield  {journal} {\bibinfo
  {journal} {Phys. Rev. X}\ }\textbf {\bibinfo {volume} {4}},\ \bibinfo {pages}
  {031052} (\bibinfo {year} {2014}{\natexlab{b}})}\BibitemShut {NoStop}%
\bibitem [{\citenamefont {Richter}\ \emph {et~al.}(1986)\citenamefont
  {Richter}, \citenamefont {Mandelberg}, \citenamefont {Kruger},\ and\
  \citenamefont {McGrath}}]{Richter1986}%
  \BibitemOpen
  \bibfield  {author} {\bibinfo {author} {\bibfnamefont {L.}~\bibnamefont
  {Richter}}, \bibinfo {author} {\bibfnamefont {H.}~\bibnamefont {Mandelberg}},
  \bibinfo {author} {\bibfnamefont {M.}~\bibnamefont {Kruger}}, \ and\ \bibinfo
  {author} {\bibfnamefont {P.}~\bibnamefont {McGrath}},\ }\href {\doibase
  10.1109/JQE.1986.1072909} {\bibfield  {journal} {\bibinfo  {journal} {IEEE J.
  Quantum Electron.}\ }\textbf {\bibinfo {volume} {22}},\ \bibinfo {pages}
  {2070} (\bibinfo {year} {1986})}\BibitemShut {NoStop}%
\bibitem [{\citenamefont {Feynman}\ \emph {et~al.}(2010)\citenamefont
  {Feynman}, \citenamefont {Hibbs},\ and\ \citenamefont {Styer}}]{Feynman_QM}%
  \BibitemOpen
  \bibfield  {author} {\bibinfo {author} {\bibfnamefont {R.~P.}\ \bibnamefont
  {Feynman}}, \bibinfo {author} {\bibfnamefont {A.~R.}\ \bibnamefont {Hibbs}},
  \ and\ \bibinfo {author} {\bibfnamefont {D.~F.}\ \bibnamefont {Styer}},\
  }\href@noop {} {\emph {\bibinfo {title} {Quantum Mechanics and Path
  Integrals: Emended Edition}}}\ (\bibinfo  {publisher} {Dover Publications},\
  \bibinfo {year} {2010})\BibitemShut {NoStop}%
\bibitem [{\citenamefont {Bender}\ and\ \citenamefont
  {Orszag}(1978)}]{Bender_Orszag}%
  \BibitemOpen
  \bibfield  {author} {\bibinfo {author} {\bibfnamefont {C.~M.}\ \bibnamefont
  {Bender}}\ and\ \bibinfo {author} {\bibfnamefont {S.~A.}\ \bibnamefont
  {Orszag}},\ }\href@noop {} {\emph {\bibinfo {title} {Advanced Mathematical
  Methods for Scientists and Engineers}}}\ (\bibinfo  {publisher} {McGraw-Hill,
  New York},\ \bibinfo {year} {1978})\BibitemShut {NoStop}%
\bibitem [{\citenamefont {Castin}\ and\ \citenamefont
  {Dum}(1996)}]{Castin_Dum}%
  \BibitemOpen
  \bibfield  {author} {\bibinfo {author} {\bibfnamefont {Y.}~\bibnamefont
  {Castin}}\ and\ \bibinfo {author} {\bibfnamefont {R.}~\bibnamefont {Dum}},\
  }\href {\doibase 10.1103/PhysRevLett.77.5315} {\bibfield  {journal} {\bibinfo
   {journal} {Phys. Rev. Lett.}\ }\textbf {\bibinfo {volume} {77}},\ \bibinfo
  {pages} {5315} (\bibinfo {year} {1996})}\BibitemShut {NoStop}%
\bibitem [{\citenamefont {Edwards}\ \emph {et~al.}(2012)\citenamefont
  {Edwards}, \citenamefont {Krygier}, \citenamefont {Seddiqi}, \citenamefont
  {Benton},\ and\ \citenamefont {Clark}}]{MarkLVM}%
  \BibitemOpen
  \bibfield  {author} {\bibinfo {author} {\bibfnamefont {M.}~\bibnamefont
  {Edwards}}, \bibinfo {author} {\bibfnamefont {M.}~\bibnamefont {Krygier}},
  \bibinfo {author} {\bibfnamefont {H.}~\bibnamefont {Seddiqi}}, \bibinfo
  {author} {\bibfnamefont {B.}~\bibnamefont {Benton}}, \ and\ \bibinfo {author}
  {\bibfnamefont {C.~W.}\ \bibnamefont {Clark}},\ }\href {\doibase
  10.1103/PhysRevE.86.056710} {\bibfield  {journal} {\bibinfo  {journal} {Phys.
  Rev. E}\ }\textbf {\bibinfo {volume} {86}},\ \bibinfo {pages} {056710}
  (\bibinfo {year} {2012})}\BibitemShut {NoStop}%
\bibitem [{\citenamefont {P\'{e}rez-Garc\'{i}a}\ \emph
  {et~al.}(1997)\citenamefont {P\'{e}rez-Garc\'{i}a}, \citenamefont {Michinel},
  \citenamefont {Cirac}, \citenamefont {Lewenstein},\ and\ \citenamefont
  {Zoller}}]{OriginalLVM}%
  \BibitemOpen
  \bibfield  {author} {\bibinfo {author} {\bibfnamefont {V.~M.}\ \bibnamefont
  {P\'{e}rez-Garc\'{i}a}}, \bibinfo {author} {\bibfnamefont {H.}~\bibnamefont
  {Michinel}}, \bibinfo {author} {\bibfnamefont {J.~I.}\ \bibnamefont {Cirac}},
  \bibinfo {author} {\bibfnamefont {M.}~\bibnamefont {Lewenstein}}, \ and\
  \bibinfo {author} {\bibfnamefont {P.}~\bibnamefont {Zoller}},\ }\href
  {\doibase 10.1103/PhysRevA.56.1424} {\bibfield  {journal} {\bibinfo
  {journal} {Phys. Rev. A}\ }\textbf {\bibinfo {volume} {56}},\ \bibinfo
  {pages} {1424} (\bibinfo {year} {1997})}\BibitemShut {NoStop}%
\bibitem [{\citenamefont {Kagan}\ \emph {et~al.}(1996)\citenamefont {Kagan},
  \citenamefont {Surkov},\ and\ \citenamefont
  {Shlyapnikov}}]{kagan_evolution_1996}%
  \BibitemOpen
  \bibfield  {author} {\bibinfo {author} {\bibfnamefont {Y.}~\bibnamefont
  {Kagan}}, \bibinfo {author} {\bibfnamefont {E.~L.}\ \bibnamefont {Surkov}}, \
  and\ \bibinfo {author} {\bibfnamefont {G.~V.}\ \bibnamefont {Shlyapnikov}},\
  }\href {\doibase 10.1103/PhysRevA.54.R1753} {\bibfield  {journal} {\bibinfo
  {journal} {Phys. Rev. A}\ }\textbf {\bibinfo {volume} {54}},\ \bibinfo
  {pages} {R1753} (\bibinfo {year} {1996})}\BibitemShut {NoStop}%
\end{thebibliography}%

\end{document}